# POLAR CORONAL PLUMES AS TORNADO-LIKE JETS

E. Tavabi,[1] S. Koutchmy,[2] and L. Golub[3]

[1] *Payame Noor University (PNU), 19395-3697, Tehran, I. R. of Iran,*
[2] *Institut d'Astrophysique de Paris, UMR 7095, CNRS and UPMC, 98 Bis Bd. Arago, 75014 Paris, France.*
[3] *Harvard-Smithsonian Center for Astrophysics, 60 Garden St., Cambridge MA 02138, USA.*

## ABSTRACT

We examine the dynamical behavior of white light polar plume structures in the inner corona that are observed from the ground during total solar eclipses, based on their EUV hot and cool emission line counterparts observed from space. EUV observations from SDO/AIA of a sequence of rapidly varying coronal hole structures are analyzed. Evidence of events showing acceleration in the 1.25 Mk line of Fe XII at 193 A is given. The structures along the plume show an outward velocity of about 140 $kms^{-1}$ that can be interpreted as an upwards propagating wave in the 304 A and 171 A lines; higher speeds are seen in 193 A (up to 1000 $kms^{-1}$). The ejection of the cold He II plasma is delayed by about 4 min in the lowest layer and more than 12 min in the highest level compared to the hot 193 A behavior. A study of the dynamics using time-slice diagrams reveals that a large amount of fast ejected material originates from below the plume, at the footpoints. The release of plasma material appears to come from a cylinder with quasi-parallel edge-enhanced walls. After the initial phase of a longitudinal acceleration, the speed substantially reduces and the ejecta disperse into the environment. Finally, the detailed temporal and spatial relationships between the cool and hot components were studied with simultaneous multi-wavelength observations, using more AIA data. The outward-propagating perturbation of the presumably magnetic walls of polar plumes supports the suggestion that Alfven waves propagate outwardly along these radially extended walls.

Corresponding author: Ehsan Tavabi etavabi@gmail.com

# 1. INTRODUCTION

### 1.0.1. *Coronal Holes*

Coronal holes (CH) are regions of low-density plasma in the solar atmosphere associated with magnetic field lines that open up freely towards interplanetary space (Zirker 1977). During solar minimum, such holes cover both the north and the south polar caps. In more active periods, narrow CHs can also be seen at all solar latitudes as first suggested in the late 1950s, when M. Waldmeier noticed long-lived low intensity regions in coronagraphic records made with the Fe XIV green line (e.g. Koutchmy 1977; Koutchmy and Bocchialini 1998; Wilhelm et al. 2011). Their more definite identification was done from soft X-rays disk images taken from sounding rockets (Vaiana et al. 1973a) and especially from the *Skylab* Apollo Telescope Mount (ATM) grazing incidence photographic X-ray telescope (Vaiana et al. 1973b). Long-lived bright points were extensively studied over the whole quiet Sun (Golub et al. 1974; Habbal & Withbroe 1981) and transient brightenings were analyzed using the soft X-ray data from the *Yohkoh* mission (Koutchmy et al. 1997). More specifically, CHs were seen to contain a large number of small, rapidly varying bright points often coinciding with the boundaries of supergranular cells that in turn may be related to the primary launching sites for the fast solar wind, and flares in bright points observed at the limb were seen to correspond to so-called surges and to eruptive ejections in *Hα* (Koutchmy et al. 1997, 2012; Moore et al. 1977; DeForest et al. 1998).

The plasma in the transition region and in the corona is in a dynamical state where the details of the temperature minimum network rosettes (Beckers 1968; Sterling 2000; Lorrain et al. 2006; Tavabi 2018) and the magnetic field are measured, and the network patches higher up are observed in *Hα* or in chromospheric lines (Baudin et al. 1996; Georgakilas et al. 2001; De Pontieu et al. 2014). The plasma flows successively along various magnetic field tubes in the form of spicules and jets, and finally along polar plumes. The coronal material extends towards the outer corona where the solar wind actually evolves but this behavior does not take into account the intermittence of the evolution. It is accepted that the fast solar wind emerges from magnetically open CHs, which are representative of a predominantly unipolar magnetic region (e.g. Koutchmy 1977; Zirker 1977; Bravo & Stewart 1996), of a rather quiet Sun; the slow solar wind originates from above the more magnetically active part of the Sun. To establish the CH origin of the fast solar wind it is necessary to find the source regions of the high velocity solar wind stream at the surface of the Sun (Krieger et al. 1973).

### 1.0.2. *Polar Plumes*

Genuine polar plumes are linear, rather straight structures seen in the EUV line at 171 A up to 1.3$R_\odot$ and in white-light during eclipses to much larger distances (see Figure 1 for a convincing correspondence). They should not be confused with the edge of streamer sheets seen in projection above the polar limbs (Bravo & Stewart 1996). In Figure 1 we illustrate the correlation between the classical white-light plumes from eclipse observations and the plumes seen in the excellent 171 A AIA/*SDO* filtergrams after using a summing of 80 consecutive images with cadence of 12 sec. Outwardly propagating fast features are visible in hot coronal lines along the plumes as illustrated in Figure 1, EUV 171 A structures with W-L structures making the ray-like features in the polar region and usually called polar plumes or polar jets. Further, we looked at the relationship with He II 304 macrospicules and polar erupting small prominence structures, as was suggested from simultaneous ground based observation performed at the feet of the plume in the cool *Hα* line (Georgakilas et al. 2001). Macrospicules were broadly defined in He II 304 A observations as long, usually pointed jets ranging in length from 5 to over 50 arcsec (Bohlin et al. 1975).

The ions in high-speed solar wind accelerate to terminal speeds of about 800 $kms^{-1}$ above the CH (Wilhelm et al. 2000, 2011). This rapid acceleration cannot be explained by hydrodynamic forces (plasma gas pressure) alone. Some other form of momentum transfer, such as forces appearing in the region due to self-excited dynamo must be at work (Lorrain et al. 2006), or other mechanisms, such as Alfven-cyclotron resonance (Kasper et al. 2008). Coronal plumes, together with other dynamic structures in the solar atmosphere, such as macrospicules and X-ray jets, are potential sources of solar wind (spicules provide 100 times more mass than what is needed to fill the corona Pneuman and Kopp (1978)) and moreover, recent studies indicate that plumes are a plausible source of the fast solar wind streams (Gabriel et al. 1994, 2009). Other investigations led to different results (e.g. Wang et al. 1998; Habbal et al. 1995; Wilhelm et al. 2000). Wang et al. (1998) reported speeds of 400-1100 $kms^{-1}$ for the outermost part of the W-L jets, whereas their centroidal velocities for the bulk of material are much lower. The contribution of polar plumes to the fast solar wind rising from the polar CH is thus still a subject of debate and controversy (e.g. Habbal 1992; Gabriel et al. 2003). More details about plumes and inter-plume regions were listed in the review of Wilhelm et al. (2011) and of Poletto (2015).

Del Zanna et al. (2003) found that bright points are seen near the plume footpoints only in the early phase of their formation. Wilhelm et al. (2010) found a strong association of plumes and jets from XRT/Hinode images and Raouafi et al. (2008) showed that X-ray jets are precursors of coronal plume formation. Studies of morphology, kinematics, dynamics and the production mechanism of a rather small-scale tornado have been recently pursued (Wedemeyer-Bohm et al. 2012) at photospheric granulation scale. The evolution as a function of height towards the corona of this type of event is still unknown. Kitiashvili et al. (2013) provided a realistic numerical simulation showing that the small jets due to the vortex at the foot of a plume could capture and stretch the background magnetic field, generate shocks and push plasma to the upper atmosphere. The magnetic field captured in the vortex causes an initial perturbation and accelerates the plasma along the helical medium to produce a spontaneous-seeming ejection in the solar atmosphere (Lorrain et al. 2006). A great number of observations and studies show that tornado-type eruptions (including surges) are composed of fine, long threads (Tavabi et al. 2015b; Chen et al. 2012). The AIA observations of polar CHs suggest that the region may be finely structured at arcsec scales in width with multiple threads of both hot and cool dynamical tornados.

### 1.0.3. *EUV Jets*

X-ray jets were discovered with the X-ray telescope onboard the Japanese *Yohkoh* mission (Yokoyama and Shibata 1995; Shibata et al. 1992). They are associated with small flares and occur above X-ray bright points (Golub et al. 1974). Moore et al. (1976, 1977) also showed that X-ray bright points could be associated with *Ha* macrospicules. Shibata et al. (1992) and Canfield et al. (1996) have reported that X-ray jets are observed to be associated with cool ejections or plasmoids, nearly simultaneously and in nearly the same direction.

There have been numerous studies predicting that twisting motions should be present in eruptive structures in the low corona. Radially-directed linear structures seen above the limb exhibit fast ejections that reach peak velocities of 20-200 $kms^{-1}$ (Tavabi et al. 2011, 2015a; De Pontieu et al. 2014). At chromospheric heights they are called spicules, and usually they exhibit a rotational motion with several turns (Tavabi et al. 2013). However, the rotational behavior seems uncommon in X-ray jets (e.g. Filippov et al. 2009).

Pariat et al. (2010) performed 3D MHD simulations of an instability taking place in a quasi-statically driven MHD system showing a jet; they found that torsional Alfven waves with upflowing helical motion are the result of reconnection of initially twisted loops with open fields. Canfield et al. (1996) suggests that the rotational motion is due to the redistribution of stored twist after reconnection with two loops having different numbers of 360° turns. Patsourakos et al. (2008) reported on observations made by the twin spacecraft *STEREO* A and B in the EUV (195 A and 171 *A)* of a helical jet. This jet twists and untwists above the limb in a polar coronal hole.

Moore et al. (2015) studied several cases of hot EUV jets in 193 *A* line, with erupting loops that are also seen as bright emission in the relatively cool 304 *A* line with darker components (absorption) seen in the 193 *A* hotter line at the base of an eruptive tornado-like event, which is also seen in our study (see Figures 2 and 3). This schema suggests that the interacting loops at the reconnection X-null point are related to the rather cool chromospheric features and with the initial plasma ejection being due to mutual interactions of highly twisted cool loops. Tornado phenomena were occasionally reported with a turn number of more than 3 full 360° turns (Shen et al. 2011). The number of turns and consequent untwisting behavior should be taken into account when interpreting the energy release and the acceleration mechanism.

Panesar et al. (2013) reported tornados as dynamical, conspicuously helical magnetic structures mainly observed as a prominence activity that the tornado was dynamically associated with the expansion of the prominences helical field and its cavity.

## 2. OBSERVATIONS

The Atmospheric Imaging Assembly (AIA) onboard the *Solar Dynamics Observatory (SDO)* (Pesnell et al. 2012) is an array of four telescopes that captures images of the Sun's atmosphere out to 1.3R☉ in ten separate EUV and UV wave bands (Title et al. 2006). The images are 4096 x 4096 square with a pixel size of 0.6" in the full spatial resolution mode, with a cadence of 12 seconds on 10 July 2010 (Figure 2) that are considered here. A full day sequence was prepared by Koutchmy et al. (2012) for the analysis of events observed at the time of the total solar eclipse of 2010, to increase the signal/noise ratio, and improve the visibility of images. The processing method is based on frame summing, radial filtering and unsharp masking.

A solid line in Figure 4 indicates the position along which the jet was ejected. We confirm the association of plumes with small-scale explosions at their feet using a movie assembled from the AIA data (movie S1). Several examples of dynamical jets were found at the pole over a time interval of ~ *24H*. They were seen at the base of much more extended plumes. To quantify these motions, we did three cuts along the axis of the selected jet, oriented roughly perpendicular to the axis as shown in Figure 4. It is difficult to measure the radial velocity of the bulk of the plasma within the jets but, given the large number of such events now being seen in the 171 *A* channel using running difference frames (Figure 5), it seems that in 193 *A* hotter line the speed is large enough (table 1) for the plasma to escape and to significantly contribute to the fast solar wind; the typical velocity is about 800 $kms^{-1}$.

Time-slice diagrams (Figure 6) of the cuts clearly show the dynamical behavior of the trapped ejected plasma inside the plume.

Based on this diagram, we suggest that the early phase of the plasma ejection, after a phase of circular motion inside the plumes, is followed by an evaporation phase (ejection) along the plume after the plasma emerges into the corona in an open-field environment, similarly to that proposed by *e.g.* Wilhelm et al. (2011). An initial ascending radial velocity is shown by the green line of the time-distance diagram in Figure 7 (result for the 304 *A* line). It shows a typical value of $140 \pm 10$ $kms^{-1}$ found for the ascending velocity. The rotational velocity is found of order of 0.02 rad/sec ($145 \pm 30$ $kms^{-1}$); the deduced period of rotation is near 5 min. This rotational velocity is deduced assuming a spiral motion of identified features along what looks like a cylindrical surface. The average outflow velocity is the same for the 171 *A* line and remains approximately constant during the ascent. Considering the velocity of the intensity front, we find for the 193 *A* line velocity of $720 \pm 10$ up to more than $1000 \pm 10$ $kms^{-1}$ during the ascent, which is in significant contrast with the apparent velocities observed in the cool 304 *A* line (Figure 7).

In order to investigate this possible scenario of a helical motion followed by evaporative ejection, we analyzed the behavior of the cool part of our 304 *A* jet in Figure 7 using the time-distance method. The most significant result is the evidence (see the right part of the diagram of Figure 7) of time-distance diagram following quasi-parabolic trajectories. Some returning flux is also seen although the innermost part of the phenomenon inhibits a definite ejection. We note that a similar result was obtained (Kayshap et al. 2013) for a polar event interpreted as a macrospicule and the associated jet, without considering the coronal counterparts. In this study, we present detailed observations of a CH jet, with a cool tornado-like structure at the feet and extended in the radial direction.

## 3. DISCUSSION

To reveal the importance of the mechanism causing tornado-like events, the events were simultaneously observed in a time sequence of data obtained at multiple wavelengths in order to understand their dynamical characteristics.

The first important aspect of this observation is the occurrence of jets as plasma ejections related to polar plumes. Raouafi and Stenborg (2014) also reported several jetlets (small jets) at the feet of plumes (see also white light observations from eclipse images, Koutchmy & Stellmacher (1976)). Jets are highly dynamic and are seemingly one of the main energetic material source for plumes. It was also noted that jet ejection along the preexisting plumes increases the plume brightness. Wilhelm et al. (2011) show that polar jets in the interplume region propagate along the plumes seen in the Large Angle Spectroscopic Coronagraph (LASCO) instrument on *SoHO* (Brueckner et al. 1995) white-light images (Figure 8), and in eclipse white-light images near the solar minimum of activity. An extended analysis of the activity in the polar regions was made by Wang et al. (1998) using the EIT EUV filtergrams and the LASCO C2 images during the *SoHO* mission with a more limited spatio-temporal resolution compared to the present study. Note that these authors called the EUV eruptive phenomenon event producing a polar W-L dynamical structure a dynamical jet.

In Figure 10 we show more details and improve the visibility of the dynamical parts, the original intensities of plumes in each image were subjected to a running subtraction using a background image obtained after summing 16 minutes of consecutive images with cadence of 12 sec which run over the whole time sequence, pixel by pixel. In order to reproduce physically meaningful intensities we added a threshold to the result of the subtraction in Figure 5.

In the temporal evolution from the 304 *A* images of the tornado, the main components (or kernels) of the event are launched from the base with a constant proper velocity having an average value of $140 \pm 40$ $kms^{-1}$ (from the analysis of Figure 7), and a rotational velocity with a similar value, of 145 $kms^{-1}$ in this cool line. This filtering process allows us to separate short-lived dynamic phenomena from rather longer life-time background plume intensity variations. The time-distance diagram along the axis of tornado (Figure 7) in 304 A line suggests that later the ejected plasma is flowing back to the Sun. This type of a suggested complete parabola motion path (e.g. Figure 7) is not seen in the 193 *A* line. The longitudinal motion observed in Figure 3 is confirmed by the enhancement appearing later (see Figure 8) above the same polar region and the same radial position, over the occulting disk of the C2 coronagraph of LASCO. Wang et al. (1998) already reported a large correlation between the Extreme-ultraviolet Imaging Telescope (EIT) jets and the long narrow white light structures in the outer corona recorded using the LASCO observations. A large maximum value of velocity, of order of 1100 $kms^{-1}$ for the leading edge of the white-light jets (the outermost part of the jet that can be visually seen and distinguished) was found.

The two-threaded 171 *A* and 193 A*l* jets traveled at an apparent velocity of 140 $kms^{-1}$ in Fe IX (171 A*l*) and up to 600 $kms^{-1}$ to 1000 $kms^{-1}$ in Fe XII (193 A*l*) images, with a median value of 720 $kms^{-1}$ (Figure 10), 304 A*l* shows speed of 140 $kms^{-1}$, similar to what is observed for 171 A*l* (Figures 3 to 5). Such values in He II line are also reported from EIS observations by Kamio et al. (2010) for macrospicules seen associated to X-ray jets.

## 4. CONCLUSION

In this study, we present detailed observations of a coronal hole jet, with a cool helical tornado-like structure at the feet and an extended hotter component in the radial direction. Our detailed analysis (Figures 9 to 11) shows that the propagation velocities are different for different line emissions and, accordingly, for different temperatures. In Figure 11 we show that at the highest level of the event the hot emissions occur first (also see Figure 10). This could suggest that the coronal heating is occurring first at higher levels where the densities are smaller.

The foot region is presumably above a well-developed 10" size network element (rosette) of chromosphere. Recently Wedemeyer-Bohm et al. (2012) and Kitiashvili et al. (2013) considered an atmospheric tornado-like phenomenon as the result of a vortex magnetic structure driven by turbulence motions. It was also called a swirl with helical motion. Only a small scale (1 Mm) region was analyzed (Wedemeyer-Bohm et al. 2012; Kitiashvili et al. 2013), while here we deal with a region one order of magnitude larger. A simple model which permits a twisted magnetic field could imply a self-sustained ring current dynamo occurring in the high photosphere (Lorrain and Koutchmy 1996; Veselovsky et al. 1999) to enhance the magnetic field, as a result of converging horizontal motion. We propose that the main steps of tornado evolution consist of three phases (see Figures 3, 9 and 10): (1) a precursor heating phase where new magnetic flux emerges to the reconnection site with heating by current sheet at the X-null point made by interacting loops, (2) the impulsive phase that begins when the cool chromospheric loops start to rotate, and (3) the main phase, when the stored magnetic energy is released by ejection and untwisting.

Table 1. Time delay and life-time in three lines and at different cuts, marked in Figure 4, as deduced from Figures 5, 6, 7 and 10.

| Spectral ranges | Upward velocity (km/s) | Time difference (min.)±0.5 | Duration (min.)±0.5 | Diameter FWHM (Mm.±0.7) |
|---|---|---|---|---|
| Chromosphere and TR, He II, 304 Å, $Log(T) \sim 4.7$ | $140 \pm 40$ Rotational velocity$\sim 0.02$ rad/sec at h=30 arcsec above the limb, The rotational period is $\sim 300$ sec. | C3 (H=50 Mm) $\Delta t_{(304,193)} = 9.5$ | C1=20 C2=12 C3=10 Average: $\sim 14$ min. | C1=14 C2=23 C3=30 |
| Fe IX, 171 Å, Quiet corona and upper transition region, $Log(T) \sim 5.8$ | $140 \pm 40$ | C2 (H=30 Mm) $\Delta t_{(304,193)} = 8$ | C1=18 C2=16 C3=12 Average: $\sim 15$ min. | C1=C2=C3=7 |
| Fe XII, XXIV, 193 Å, Coronal plasma, $Log(T) \sim 6.1$ & $7.3$ | $\sim 720 \pm 40$ to $> 1000$ | C1 (H=10 Mm) $\Delta t_{(304,193)} = 3.5$ | C1=19 C2=21 C3=24 Average: $\sim 21$ min. | C1=C2=C3=7 |

\* C1=10, C2=30 and C3=50 Mm correspond to the different levels that were remarked in Figure 4.

The observation of a time difference between the hot linear jet-like feature and the ensuing tornado-like cool event suggests that the tornado helical formation is closely related to the precursor hot component that occurs when a release of energy occurs with the acceleration inside the tornado plasma vortex near the reconnection site. The SDO/AIA temporal and spatial resolution allows us to resolve some uncertainty concerning the relationship of hot line ejections and the cold tornado events at their footpoints. The detailed analysis of the behavior of different components of our event strongly supports the rather new suggestion that polar plumes (e.g. see Figure 1) are made of simple cylindrical structures of 10" to 25" diameter with an expansion in time, height and width. A more detailed comparison with the sub- structure and the behavior of eclipse white-light polar-plumes is needed (work in progress) to understand these structures. The forthcoming launch by ESA of a novel coronagraph system (PROBA3 mission) permitting a total solar eclipse in space for several hours (Lamy et al. 2008) is another opportunity for making progress on this very old and classical question of explaining magnetic solar coronal polar plumes.

Acknowledgments We thank the referee for the constructive comments, and we are indebted to Robert Erdelyi and Christopher Nelson for useful discussions and remarks. We acknowledge the $AIA/SDO$ consortium for the easy access of calibrated data. The AIA data are courtesy of the SDO mission (NASA). LG was supported in this work by a contract for AIA from Lockheed Martin (LMSAL). This study was made possible through the efforts of the LASCO experiment of $SoHO$ mission (ESA-NASA) team,

E.T. are also grateful to the Iran National Foundation INSF, and of the French Institut d'Astrophysique de Paris-CNRS and UPMC provides a frame for this work.

## REFERENCES

Baudin, F. Bocchialini, K. and Koutchmy, S. 1996, A&A, 314, L9-L12

Beckers, J.M. 1968, Solar Phys., 5, 309 Bohlin, J. D., Vogel, S. N., Purcell, J. D., Sheeley, Jr., N. R., Tousey, R., & Vanhoosier, M. E. 1975, ApJL, 197, L133

Brueckner, G. E., et al. 1995, Solar Phys, 162, 357.

Bravo, S., & Stewart, G. A. 1996, GeoRL, 23, 3271

Canfield, R. C., Peardon, K. P., Leka, K. D., Shibata, K., Yokoyama, T., and Shimojo, M. 1996, ApJ, 464, 1016

Chen, H.-D., Zhang, J., and Ma, S.-L. 2012, RAA, 12, 573

K. Chandrashekhar, A. Bemporad, D. Banerjee, et al., 2014, A&A 561, 104. doi:10.1051/0004-6361/201321213 De Pontieu, B., Rouppe van der Voort, L., McIntosh, S. W., et al. 2014, Science, 346, 1255732

DeForest, C.E., Hoeksema, J. T., Gurman, J. B., Thompson, B.J., Plunkett, S. P., Howard, R., Harrison, R.C. and Hassler, D.M. 1997, Solar Phys, 175, 393

DeForest, C.E., Gurman, J.B. 1998, ApJ, 501, L217 Del Zanna, L., Hood, A.W. and Longbottom, A.W. 1997, A&A, 318, 963

Del Zanna, L., Bromage, B.J.I. and Mason, H.E. 2003, A&A, 398, 743

Filippov, B., Koutchmy, S., & Vilinga, J. 2007, A&A, 464, 1119

Filippov, B., Golub, L., Koutchmy, S.: 2009, Solar Phys. 254, 259. DOI: 10.1007/s11207-008-9305-6

Gabriel, A. H., Bely-Dubau, F. and Lemaire, P. 2003, ApJ, 589, 623

Gabriel, A.H., Abbo, L., Bely-Dubau, F., Llebaria, A., Antonucci, E. 2005, ApJ, 635, L185 Gabriel, A., Bely-Dubau, F., Tison, E. and Wilhelm, K. 2009, ApJ, 700, 551

Georgakilas, A. A., Koutchmy, S. and Christopoulou, E. B. 2001 A&A, 370, 273

Golub, L. Krieger, A.S. Silk, J.K. Timothy, A.F. and Vaiana, G.S. 1974, ApJ, 189, L93 Habbal, S. R., & Withbroe, G. L. 1981, Sol. Phys., 69, 77 Habbal, S. R. 1992, Annales Geophysicae, 10, 34 Habbal, S.R., Esser, R., Guhathakurta, M., Fisher, R.R. 1995, Geophys Res Lett., 22, 1465 Pesnell, W. D., Thompson, B. J., & Chamberlin, P. C. 2012, SoPh, 275, 3.

Pneuman, G.W., Kopp, R.A.: 1978, Solar Phys. 57, 49

Kasper, Lazarus & Gary, 2008. Phys. Rev. Letters, 101, 26

Kamio, S., Curdt, W., Teriaca, L., et al., 2010, A&A, 510, 1, doi:10.1051/0004-6361/200913269 Kayshap, P.; Srivastava, A. K.; Murawski, K.; Tripathi, D.; 2013;ApJ; 770, 3

Koutchmy, S. 1977, Solar Phys., 51, 399 Koutchmy, S. Hara, H. Shibata, K. Suematsu, Y. and Reardon, K., 1998, in ''Observational Plasma Astrophysics: Five Years of Yohkoh and Beyond", T. Watanabe et al. (eds.), Kluwer Acad. Publ., 87-94

Kitiashvili, I. N., Kosovichev, A. G., Lele, S. K., Mansour, N. N., and Wray, A. A. 2013, The Astrophysical Journal, 770, 37

Koutchmy, S. 1971, A and A, 13, 79

Koutchmy, S., & Stellmacher, G. 1976, Sol. Phys., 49, 253

Koutchmy, S., Hara, H., Suematsu, Y., and Reardon, K. 1997, A&A, 320, L33

Koutchmy, S. and Bocchialini, K. 1998, ESA, SP-421, 51

Koutchmy, S., Bazin, C., Berghmans, D., De Groof, A., Druckmller, M., Tavabi, E., Engell, A., Filippov, B., Golub, L., Lamy, Ph., and 6 coauthors 2012, EAS Publications Series, 55, 223

Krieger, A. S., Timothy, A. F., & Roelof, E. C. 1973, Solar Physics, 29, 505

Lamy, P., Vives, S., Dam, L., Koutchmy, S., "New perspectives in solar coronagraphy offered by formation flying: from PROBA-3 to Cosmic Vision", Proc. SPIE 7010 (2008)

Lorrain, P. and Koutchmy, S. 1996, Sol. Phys., 165, L115

Lorrain, P., Lorrain, F, and Houle S. 2006 Magneto-fluid Dynamics: Fundamentals and Case Studies of Natural Phenomena (New York, Springer), p. 319 Madjarska M. S., Vanninathan K., Doyle J. G., 2011, A&A, 532, L1

Moore, R.L., Tang, F., Bohlin, J.D., Golub, L., 1976, Bull. Am. Astron. Soc. 8, 333

Moore, R. L., Tang, F., Bohlin, J. D., Golub, L. 1977, ApJ, 218, 286

Moore, R. L., Sterling, A. C., and Falconer, D. A. 2015, ApJ, 806, 11

Newkirk, G. Jr. and Harvey, J. 1968, 3, 321 Panesar, N. K., Innes, D. E., Tiwari, S. K., & Low, B. C. 2013, A&A, 549, A105

Patsourakos, S., Pariat, E., Vourlidas, A., Antiochos, S. K., and Wuelser, J. P. 2008, ApJ, 680, L73 Pariat, E., Antiochos, S. K., and DeVore, C. R. 2009, ApJ, 691, 61. 2010, ApJ, 714, 1762 Poletto, G. 2015, LRSP, 12, 7

Raouafi, N.E., Petrie, G.J.D., Norton, A.A., Henney, C.J., Solanki, S.K. 2008, ApJ, 682, L137 Raouafi, N.-E., and Stenborg, G. 2014, ApJ, 787, 118

Shen, Y., Liu, Y., Su, J. and Ibrahim, A., ApJL 735, 43 (2011). doi:10.1088/2041-8205/735/2/L43 Shibata, K. Nitta, N. Strong, K.T., et al., 1994,ApJL 431, 51. doi:10.1086/187470 Shibata, K., et al. 1992, PASJ, 59, 771

Sterling, A. C. 2000, SoPh, 196, 79 Tavabi, E., Koutchmy, S., and Ajabshirizadeh, A. 2011, Adv. Space Res., 47, 2019

Tavabi, E., Koutchmy, S., and Ajabshirizadeh, A. 2013, Solar Phys. 283, 187. DOI: 10.1007/s11207-012- 0011-z

Tavabi, E., Koutchmy, S., Ajabshirizadeh, A.,
   Ahangarzadeh Maralani, A.R., Zeighami, S.: 2015, A&A. 573, A4. DOI: 10.1051/0004-6361/201423385 Tavabi, E., Koutchmy, S., and Golub, L. 2015, SoPh, 290, 2871

Tavabi, E. 2018, MNRAS, 476, 868

Title A. M., Hoeksema, J. T., Schrijver, C. J., & The Aia Team 2006, COSPAR Plenary Meeting, 36th COSPAR Scientific Assembly, CD ROM 2600 Veselovsky, I. S., Zhukov, A. N., Koutchmy, S., Delanne, C., Delaboudinire, J.-P., 8th SOHO Workshop, 1999, SP 446. Edited by J.-C. Vial and B. Kaldeich-Schmann., p.675 Vaiana, Krieger and Timothy, 1973. Sol. Phys. 32, 81. Vaiana, G. S., Davis, J. M., Giacconi, R., et al 1973, ApJ, 185, L47

Wang, Y.-M., Sheeley, Jr., N. R., Socker, D. G., et al. 1998,
   ApJ, 508, 899

Wedemeyer-Bohm, S., Scullion, E., Steiner, O., et al. 2012,
   Nature, 486, 505

Wilhelm, K., Dammasch, I.E., Marsch, E. and Hassler,
   D.M. 2000, A&A, 353, 749

Wilhelm, K., Dwivedi, B.N., Curdt, W. 2010, Astrophys Space Sci Proc, pp 454-458, Springer, Berlin Heidelberg Wilhelm, K., Abbo, L., Auchre, F., Barbey, N., Feng, L., Gabriel, A. H., Giordano, S., Imada, S., Llebaria, A., Matthaeus, W. H., and 5 coauthors 2011, A and A Rev., 19, 70

Yokoyama, T., & Shibata, K. 1995, Nature, 375, 42 Zhang, Q.M. and Ji, H.S., A&A 2014, 561, 134, doi:10.1051/0004-6361/201322616 Zirker, J. B. 1977, Coronal Holes and High Speed Wind Streams (Colorado Associated University Press: Boulder, Colorado)

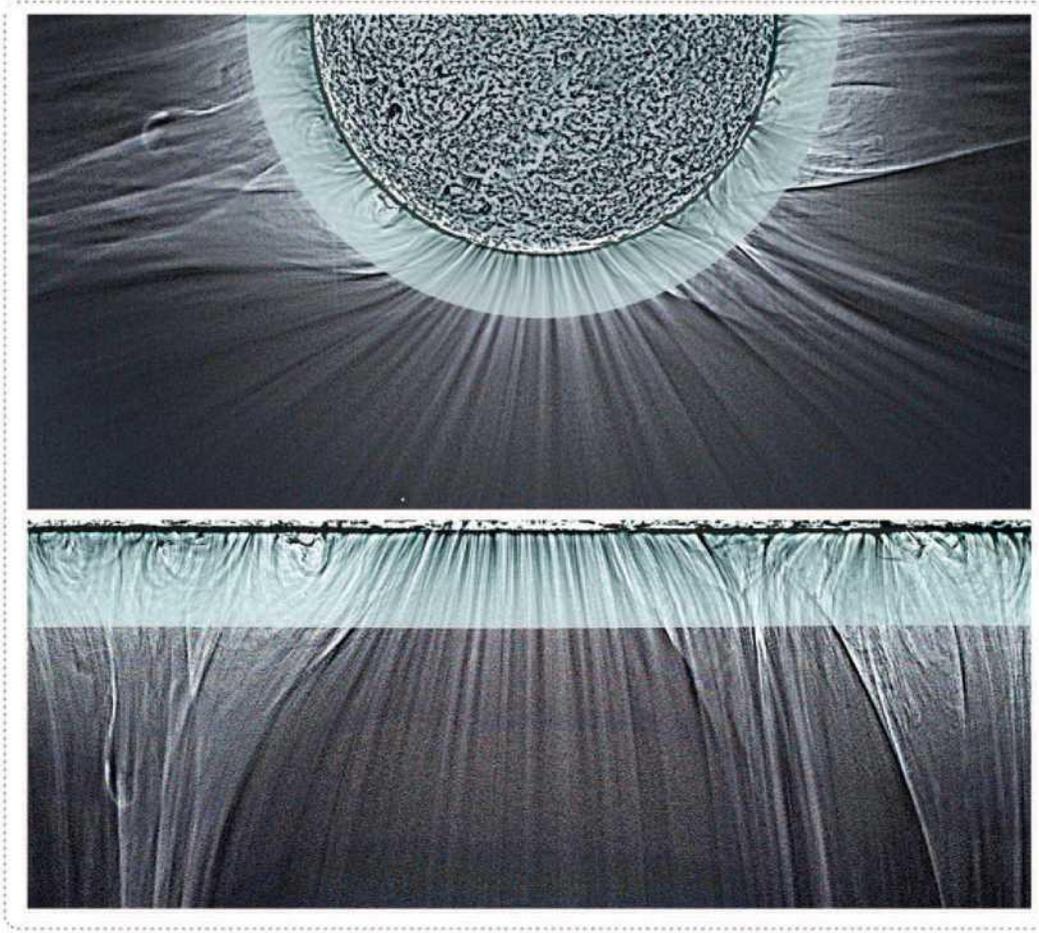

**Figure 1.** Composite white-light image of 11 July 2010 total solar eclipse taken at 20:14 UT on Easter Island in the southeastern Pacific Ocean (external part) and SDO/AIA observations in EUV 171 A line (inner part), and in polar coordinates (bottom). The AIA image corresponds to 150 frames with a 12 sec cadence after summing intensities between 20:00 to 20:30 UT and unsharp masking; intensities are in logarithmic scale. Original eclipse white-light image courtesy of Jean- Marc Lecleire.

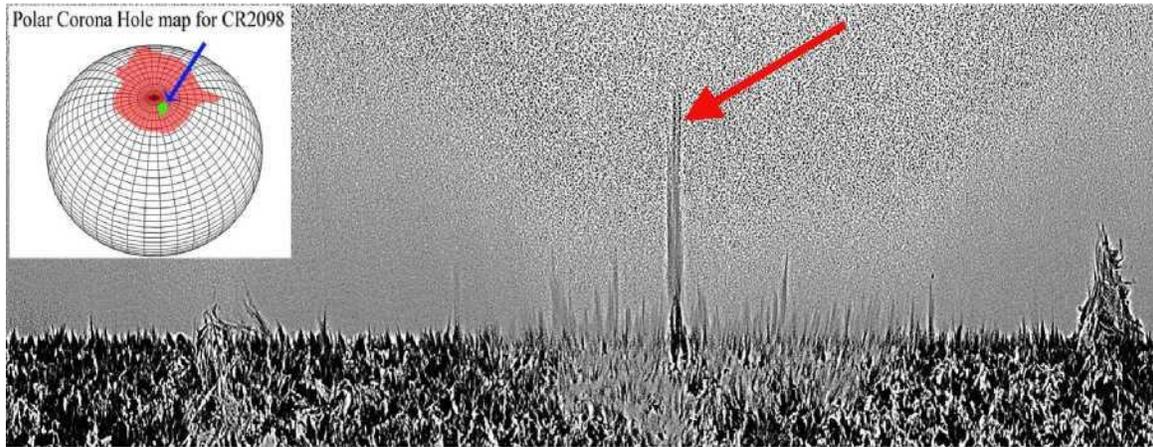

**Figure 2.** Mapping in polar coordinates *(R –θ)* of the limb around the north CH observations of July 10, 2010 (the day before the total solar eclipse), as observed in 304 *A* after summing of 80 frames with a 12 sec cadence which corresponds to 16 min between 14:28:52 to 14:44:40 UT, with unsharp masking. Negative intensities are in logarithmic scale. The longest jet is seen near the center of the figure (indicated by the red arrow), showing the tornado context. The position of the selected ejection is shown by the arrow at the top left in the north polar CH panel; it corresponds to the 2098 Carrington rotation.

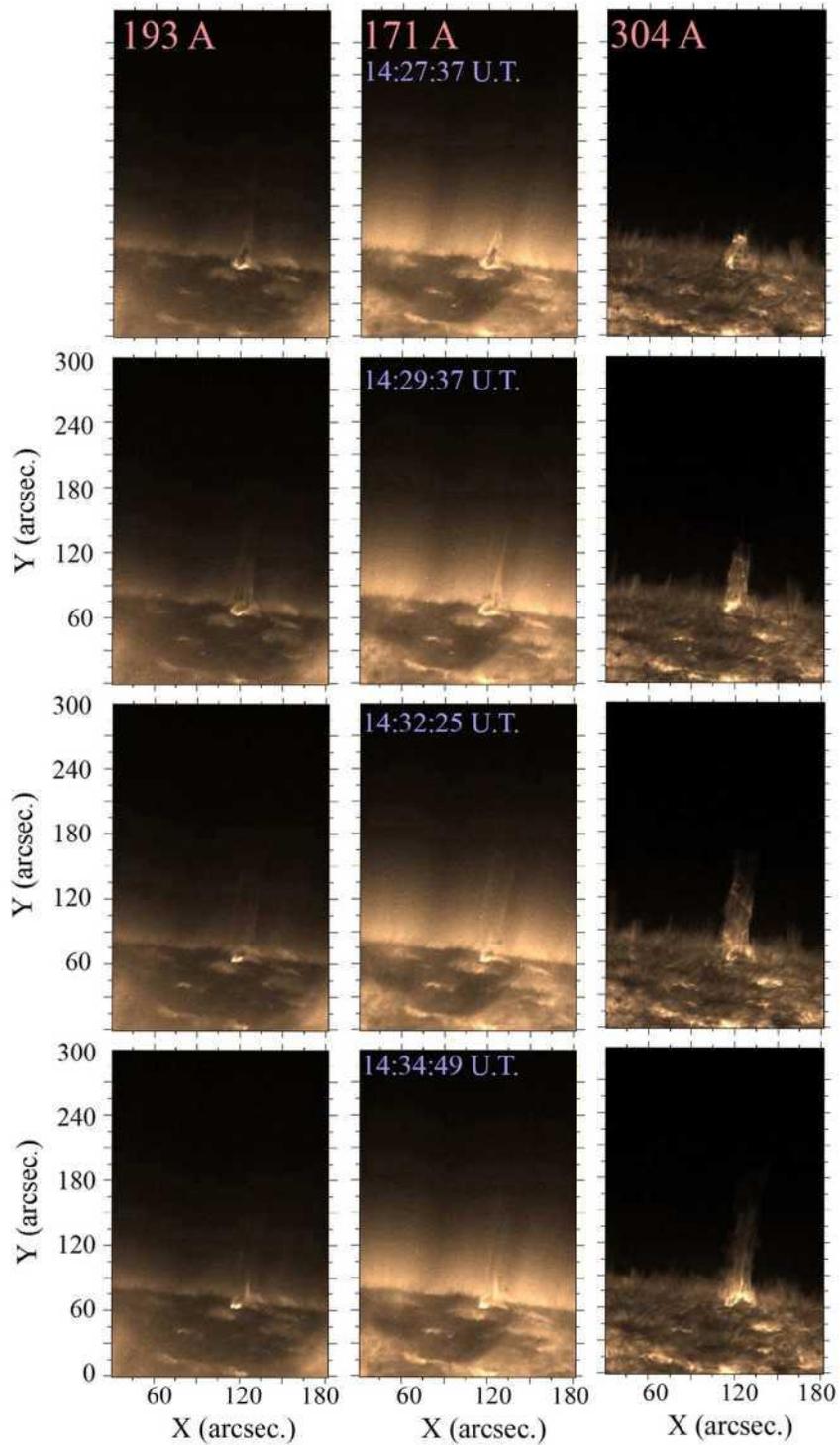

**Figure 3.** Snapshots of AIA 193, 171 and 304 *A* showing the time evolution of the tornado-like phenomenon at the base of the selected jet in 304 A and what is recorded in the coronal lines. A movie showing the temporal evolution is available in the online edition (the untwisting motion in the online available movie (S1) is more obvious when we manually run the movie at a higher speed than the default speed). Snapshot at 14:32 UT in 304 A line shows the main axis of tornado with different cuts used to study the longitudinal motion.

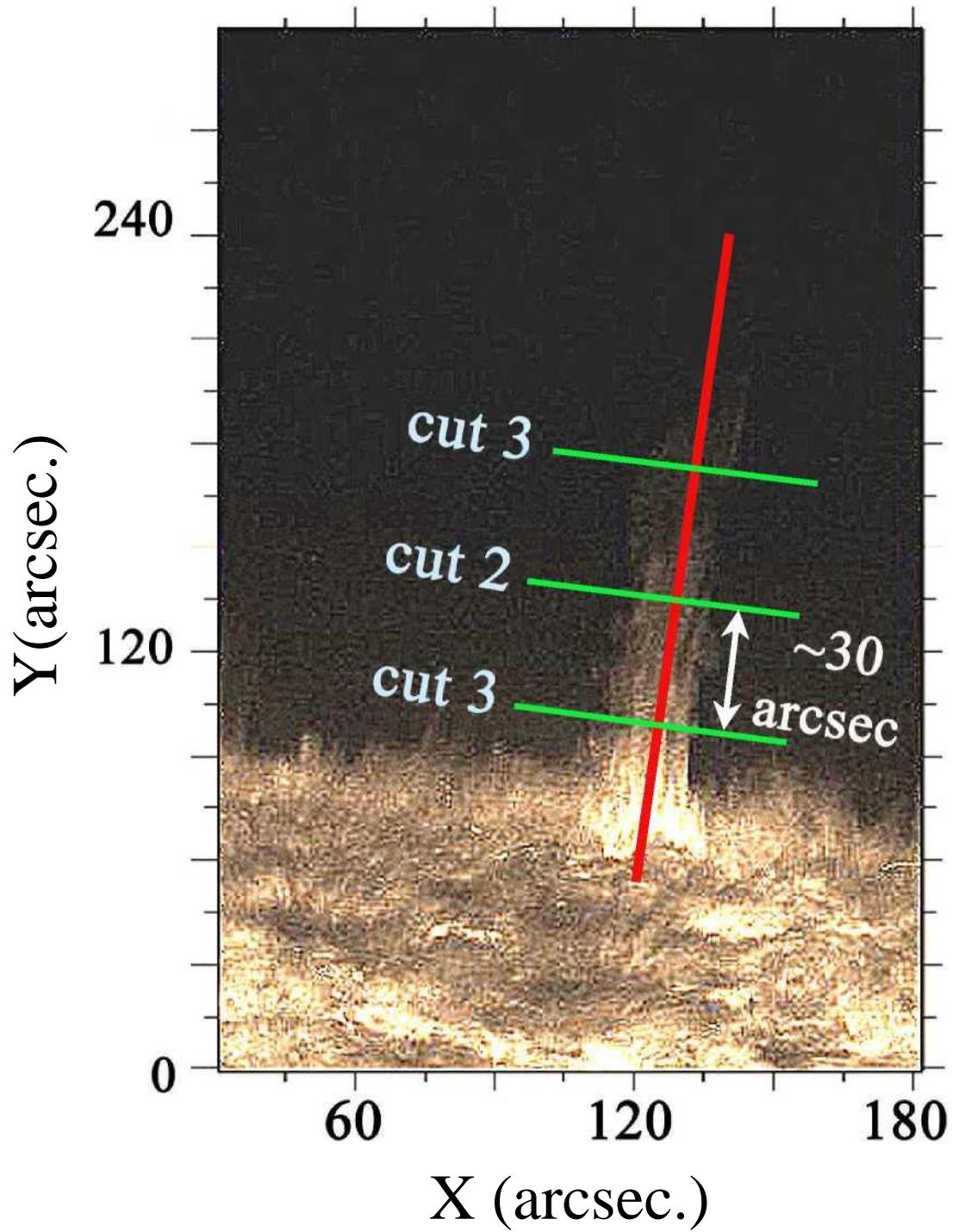

**Figure 4.** Last snapshot of Figure 3 in 304 A line to show the main axis of tornado with different cuts used to study the longitudinal motion.

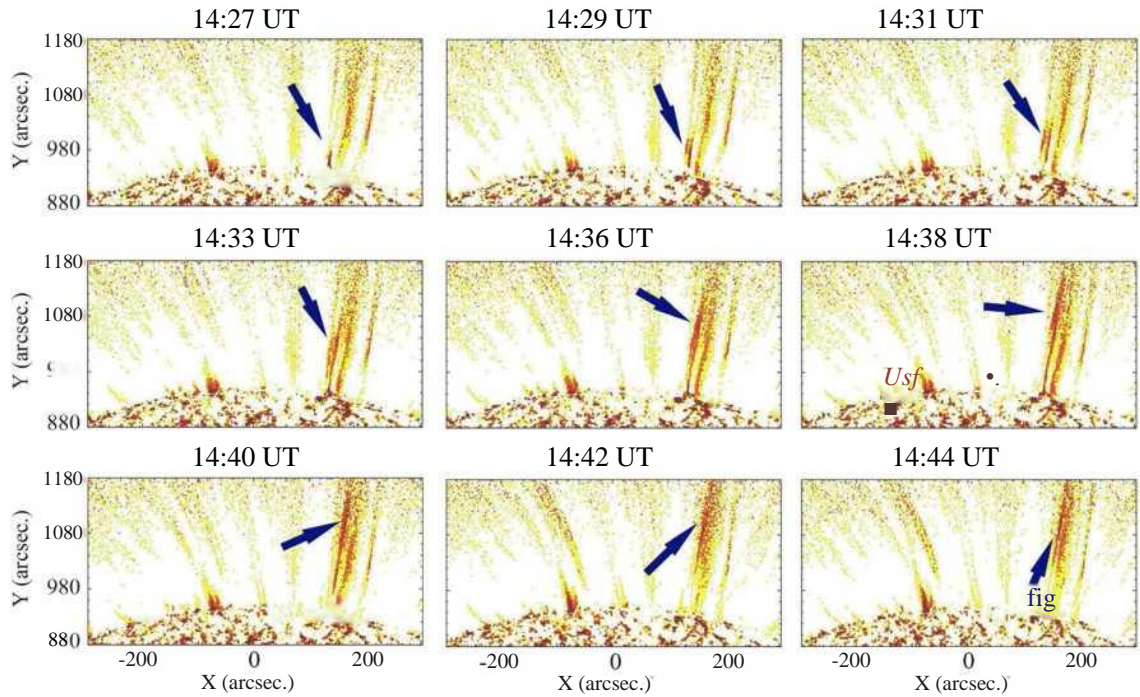

**Figure 5.** Time variation showing the acceleration inside a typical coronal plume structure. Filtered difference successive images in AIA 171 A. A movie is available in the online material (movie S2).

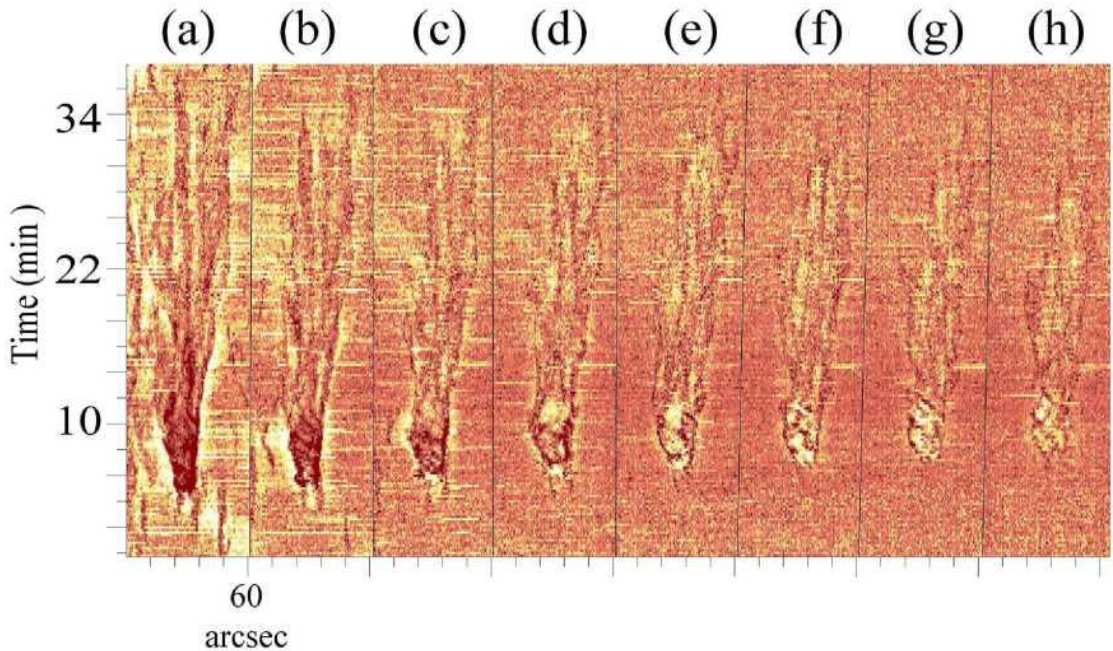

**Figure 6.** Time-slices of different cuts in the 304 $A$ filtergrams as shown in Figure 4 with evidence of untwisting plasma for 8 levels with 10 arcsec differences in heights between adjacent layers from cut 1 to cut 3 (Figure 4) to see more continuity; some expansion with time appears.

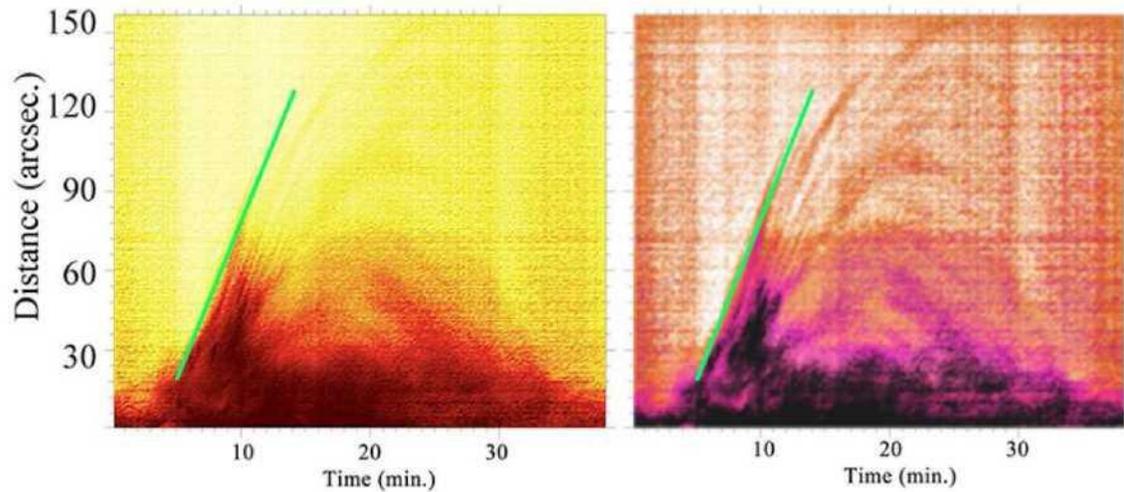

**Figure 7.** Intensity flux time-distance diagrams of the dynamical part of the jet of width 6" in 304 A. Cuts in the perpendicular direction of the jet main axis of the event shown in Figure 4 with a red line. Left: original display. Right: after increasing the visibility in the background using an unsharp masking filter with evidence of plasma clouds following a quasi-parabolic path in time-distance diagram after 15 min is demonstrated. Vertical and horizontal lines of the graph are artifacts and they are seen only after greatly increasing the contrast of frames. The apparent speed is obtained from the slope shown as a green diagonal line, the green line indicates an 140 $kms^{-1}$ trajectory.

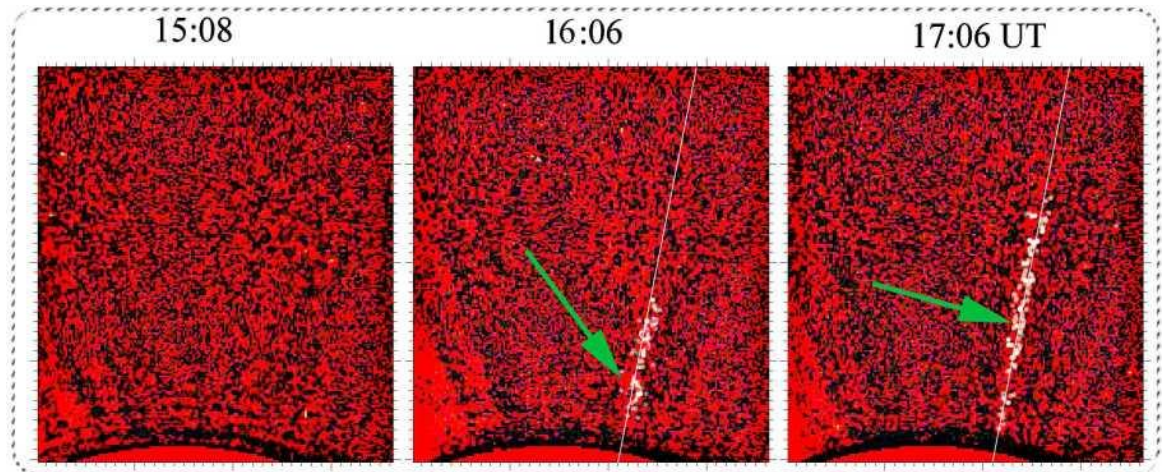

**Figure 8.** Unsharp-masked LASCO C2 sub-frame images corresponding to the tornado of Figure 3; the green arrows indicate the signature of the tornado in the outer corona, the long blue line shows the trajectory of the jet. Note that at 15:08 no bright feature is observed as the jet material has not yet reached these heights.

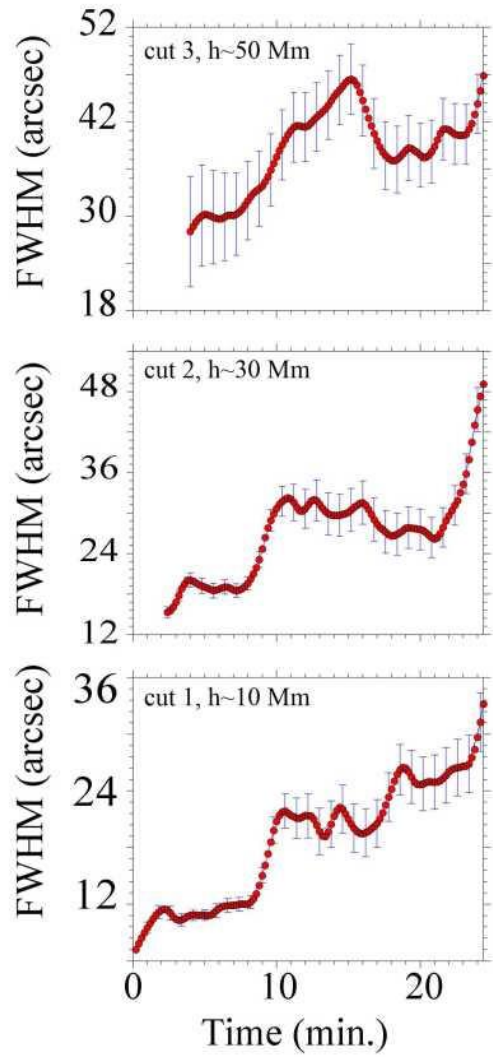

**Figure 9.** Average FWHM of the untwisting jet in 304 *A,* see Figure 4, averaged over 5 pixels (3 arcsec.) spatially and 5 steps (1 min) in time.

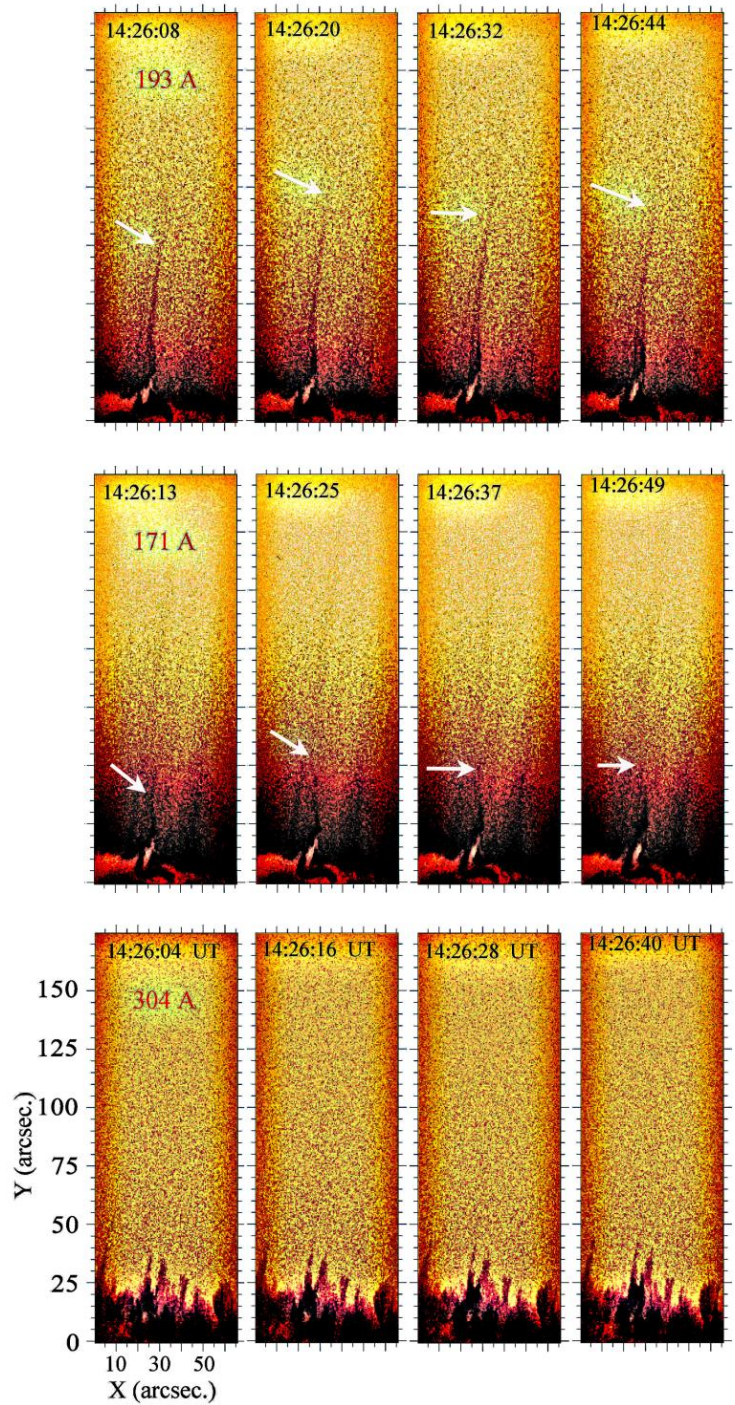

**Figure 10.** Simultaneous fast successive snapshots during the initial phase of the ejection in different filtergrams. The white arrows point to the thin linear precursor jet of the tornado; it is seen in the hotter frames at 193 A and weaker in 171 A. These frames are taken several minutes before the start of the tornado 304 A event listed in, see Table 1 and future analyzed in Figure 11.

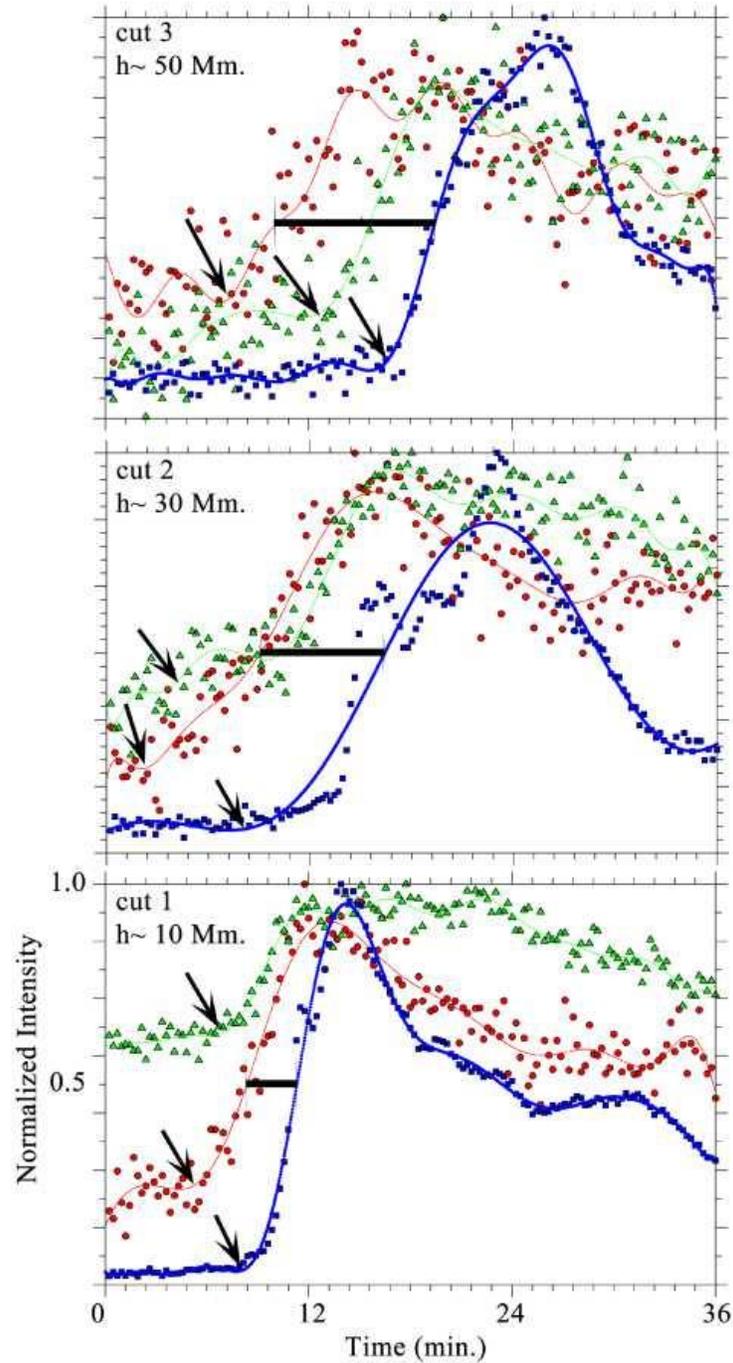

**Figure 11.** Normalized intensity evolution for the three cuts along the jet which are indicated in Figure 4, the red dots are for 193 A, the green triangles for 171 A, and the blue squares for 304 A The lines show the best Fourier fitted data. The short black arrows show the beginning phase of the sharp ascent of averaged intensities for different lines. Heights of cuts are shown in Figure 4. The thick black horizontal lines show the time difference of the ejection recorded in the cool TR (304 A) line compared to the hot coronal line (193 A).